\documentclass{article}

\usepackage{arxiv}

\usepackage[utf8]{inputenc} 
\usepackage[T1]{fontenc}    
\usepackage{hyperref}       
\usepackage{url}            
\usepackage{booktabs}       
\usepackage{amsfonts}       
\usepackage{nicefrac}       
\usepackage{microtype}      
\usepackage{lipsum}
\usepackage{lscape}
\usepackage[authoryear, sort]{natbib}

\usepackage{morefloats}
\usepackage{color}
\usepackage{multirow}
\usepackage{latexsym}
\usepackage{amsmath,amssymb,amsthm,graphicx}
\usepackage{dsfont}
\usepackage{enumitem}

\newtheoremstyle{thm}
{9pt}
{9pt}
{\itshape}
{}
{\bfseries}
{.}
{ }
{}
\theoremstyle{thm}

\newtheorem{theorem}{Theorem}[section]
\newtheorem{lemma}[theorem]{Lemma}
\newtheorem{corollary}[theorem]{Corollary}

\newtheoremstyle{def}
{9pt}
{9pt}
{}
{}
{\bfseries}
{.}
{ }
{}
\theoremstyle{def}

\newtheorem{remark}[theorem]{Remark}

\newenvironment{prf}{\textbf{\emph{Proof.}}}{\qed}

\newcommand{\R}{\mathbb{R}} 
\newcommand{\N}{\mathbb{N}} 
\newcommand{\E}{\mathbb{E}} 

\renewcommand{\footnoterule}{%
	\kern -3.5pt
	\hrule width \textwidth height 1pt
	\kern 3.5pt
}

\makeatletter
\def\blfootnote{\xdef\@thefnmark{}\@footnotetext}
\makeatother

\title{New weighted $L^2$-type tests for the inverse Gaussian distribution}

\author{J.S. Allison\\
School of Mathematical and Statistical Sciences,\\ North-West University,\\ South Africa. \\
\href{mailto:James.Allison@nwu.ac.za}{James.Allison@nwu.ac.za}\\
\And S. Betsch\\
Institute of Stochastics,\\ Karlsruhe Institute of Technology (KIT),\\ Germany.\\
\href{mailto:Steffen.Betsch@kit.edu}{Steffen.Betsch@kit.edu}\\
\And  B. Ebner\\
Institute of Stochastics,\\ Karlsruhe Institute of Technology (KIT),\\ Germany.\\
\href{mailto:Bruno.Ebner@kit.edu}{Bruno.Ebner@kit.edu}\\
\And I.J.H. Visagie\\
School of Mathematical and Statistical Sciences,\\ North-West University,\\ South Africa.\\
\href{mailto:Jaco.Visagie@nwu.ac.za}{Jaco.Visagie@nwu.ac.za}\\
}

\begin{document}

\date{\today}
\maketitle


\begin{abstract}
 We propose a new class of goodness-of-fit tests for the inverse Gaussian distribution. The proposed tests are weighted $L^2$-type tests depending on a tuning parameter. We develop the asymptotic theory under the null hypothesis and under a broad class of alternative distributions. These results are used to show that the parametric bootstrap procedure, which we employ to implement the test, is asymptotically valid and that the whole test procedure is consistent. A comparative simulation study for finite sample sizes shows that the new procedure is competitive to classical and recent tests, outperforming these other methods almost uniformly over a large set of alternative distributions. The use of the newly proposed test is illustrated with two observed data sets.
\end{abstract}

\section{Introduction}
\label{Section introduction}
The inverse Gaussian distribution (also known as the Wald distribution) was first heuristically observed by \cite{B:1900}, and derived by \cite{S:1915} as the distribution of the first passage time of Brownian motion with drift, see \cite{S:1993} for a historical summary. In the statistical literature the usual parametrization of the inverse Gaussian law $\mbox{IG}(\mu,\lambda)$, $\mu,\lambda>0$, follows the representation of \cite{T:1957a,T:1957b}, namely the density is given by
\begin{equation}\label{density}
	f(x; \mu, \lambda) = \sqrt{\frac{\lambda}{2 \pi x^3}} \, \exp\left(- \frac{\lambda(x-\mu)^2}{2\mu^2x}\right),\quad x>0,
\end{equation}
and $f(x; \mu, \lambda) = 0$ for $x \leq 0$. Applications involving the inverse Gaussian family of distributions, $\mathcal{IG} = \{ \mbox{IG}(\mu, \lambda) \, : \, \mu, \lambda > 0 \}$, cover topics such as stock market prices, biology, hydrology, reliability, L\'{e}vy processes, and generalized linear models, as witnessed by the monographs specially dedicated to this family of distributions, see \cite{CF:1989}, \cite{S:1993}, and \cite{S:1999}.

The first step in serious statistical inference with this distribution is to test the fit of data to the family $\mathcal{IG}$. To be specific, let $X, X_1, X_2, \dotso$ be positive independent and identically distributed (iid.) random variables defined on a common probability space $(\Omega, \mathcal{A}, \mathbb{P})$. Writing $\mathbb{P}^X$ for the distribution of $X$, we intend to test the composite hypothesis
\begin{equation} \label{H0}
	H_0 : \mathbb{P}^X \in \mathcal{IG}
\end{equation}
against general alternatives. This testing problem has been considered in the statistical literature. The methods by \cite{BG:2015} and \cite{MKC:2001} are based on a characterization of the $\mathcal{IG}$ family by an independence property, \cite{D:2001} uses a connection to the so called Random Walk distribution, and \cite{ND:2003} propose exact tests based on the empirical distribution function of transformations characterising the inverse Gaussian law, which are commented and corrected in \cite{GO:2005}. \cite{HK:2002} use a differential equation that characterizes the Laplace transform of the $\mathcal{IG}$ family as well as a $L^2$-distance test, both using the empirical Laplace transform. \cite{KK:2012} use an empirical version of the standardized form of the cumulant generating function, \cite{Vetal:2011} propose an empirical likelihood test based on densities for $\mathcal{IG}$, and \cite{VG:2015} consider a variance ratio test of fit. Finally, \cite{KL:2014b} tackle the testing problem for the generalized inverse Gaussian family exploiting the ULAN property in connection to Le Cam theory. Although the testing problem is derived for a wider class of distributions, it still applies for testing (\ref{H0}) when it is restricted to the special case $p_0=-\frac12$ in the authors' notation, see Section 4 of the cited article. A comparative simulation study is provided by \cite{AN:2017}.

It is a common approach to exploit distributional characterizations to propose goodness-of-fit testing procedures, for an overview see \cite{N:2017}. As evidenced by the list above there are numerous characterizations of the inverse Gaussian distribution, including characterizing properties based on independence, constancy of regression of suitable functions on the sum of identically distributed random variables, random continued fractions, or on the relation between $\E [X]$ and $\E [X^{-1}]$. For details see \cite{S:1993}, Section 3, and \cite{KL:2014a}, Section 2.5, and for an introduction to characterizations for other distributions, see \cite{A:2017} and \cite{KLR:1973}. A very recent characterization identity is given in Example 5.10 by \cite{BE:2019:1}, which reads as follows.
\begin{theorem} \label{Theorem characterization}
	Let $X : \Omega \to (0, \infty)$ be a random variable with distribution function $F$, $\E [X] < \infty$ and $\E\left[ X^{-1} \right] < \infty$. Then $X$ has the inverse Gaussian distribution $\mbox{IG}(1, \varphi)$ if, and only if,
	\begin{equation}\label{charaeq}
		\E \left[ \frac{1}{2} \left( \varphi + \frac{3}{X} - \frac{\varphi}{X^2} \right) \min \{ X, t \} \right]
		= F(t), \quad t > 0.
	\end{equation}
\end{theorem}

Note that $X \sim$ $\mbox{IG}(\mu, \lambda)$ if, and only if, $\frac{X}{\mu} \sim$ $\mbox{IG}(1,\lambda/\mu)$ and therefore the family $\mathcal{IG}$ is closed under scale transformations. Since the characterization is directly related to the theory of Stein characterizations [for details on Stein operators, see \cite{LRS:2017}], we refer to the corresponding characterization of the generalized inverse Gaussian distribution, see \cite{KL:2014a}, Theorem 3.2, and to the connection with the Stein operator for the special case $(p,a,b) = (-1/2, \lambda / \mu^2, \lambda)$, using the authors' notation.

Our novel testing procedure is motivated by Theorem \ref{Theorem characterization}: We estimate both sides of (\ref{charaeq}) by their empirical counterparts, then calculate the weighted $L^2$-distance of the difference. We choose this distance since $L^2$-type statistics are widely used in goodness-of-fit testing, see \cite{BEH:2016}. In this spirit, we propose the statistics
\begin{equation*}
	T_{n} = n \int_0^{\infty} \left| \frac{1}{2n} \sum\limits_{j = 1}^{n} \left( \widehat{\varphi}_n+\frac3{Y_{n,j}}-\frac{\widehat{\varphi}_n}{Y_{n,j}^2} \right) \min\{ Y_{n,j}, t \} - \frac{1}{n} \sum\limits_{j = 1}^{n} \mathds{1}\{Y_{n,j} \leq t\} \right|^2 w(t) \, \mathrm{d}t
\end{equation*}
with $Y_{n,j} = \frac{X_j}{\widehat{\mu}_n}$, where $\widehat{\mu}_n, \widehat{\lambda}_n,$ are consistent estimators of $\mu, \lambda$ and $\widehat{\varphi}_n = \widehat{\lambda}_n / \widehat{\mu}_n$. The function $w(t)$ is a positive, continuous weight on $(0, \infty)$ with
\begin{equation}\label{weightcond}
	\int_0^\infty (t^2 + 1) \, w(t) \, \mathrm{d}t < \infty \quad \text{and} \quad n \int_0^\infty \Big| w\big( \widehat{\lambda}_n^{-1} s \big) - w(s) \Big|^3 \big( w(s) \big)^{-2} \, \mathrm{d}s \stackrel{\mathbb{P}}{\longrightarrow} 0,
\end{equation}
where $\stackrel{\mathbb{P}}{\longrightarrow}$ denotes convergence in probability (as $n \to \infty$). Note that, when suitable weight functions are chosen, we have numerically stable versions for the calculation of $T_{n}$ avoiding numerical integration, see Section \ref{Subsection computational form}. In particular, we use the weights
\begin{align*}
	w_a(t) = e^{- a t} \quad \text{and} \quad \widetilde{w}_a(t) = e^{- a t^2}, \quad t > 0,
\end{align*}
with a positive tuning parameter $a > 0$. Both satisfy the conditions in (\ref{weightcond}). A proof of this fact is given by \cite{BE:2019:2} for the first weight function, and the argument for the second weight is very similar, so we do not discuss it here. Since $\mathcal{IG}$ is scale invariant, the test should reflect this property. Thus we only consider scale equivariant estimators $\widehat{\mu}_n$ of $\mu$, i.e. we have
\begin{equation*}
	\widehat{\mu}_n(\beta X_1, \ldots, \beta X_n) = \beta \, \widehat{\mu}_n(X_1, \ldots, X_n), \quad \beta > 0.
\end{equation*}
Likewise, let $\widehat{\lambda}_n$ be an estimator of $\lambda$ which is scale equivariant and thus ensures that $\widehat{\varphi}_n$ is scale invariant, that is,
\begin{equation*}
	\widehat{\varphi}_n(\beta X_1, \ldots, \beta X_n) = \widehat{\varphi}_n(X_1, \ldots, X_n), \quad \beta > 0.
\end{equation*}
With this type of estimators it is straightforward to show that $T_{n}$ is invariant under scale transformations of the data, as it depends on (the scale invariant) $Y_{n,j}$, $j = 1, \ldots, n$, and $\widehat{\varphi}_n$ only. Rejection of $H_0$ in (\ref{H0}) is for large values of $T_{n}$.

The paper is structured as follows. In Section \ref{Section estimators} we present two estimation procedures in conjunction with the $\mathcal{IG}$-family as well as asymptotic representations of the estimators needed in the subsequent theory. Section \ref{Section asymH0} gives theoretical derivations of asymptotic results under the null hypothesis, and Section \ref{Section contiguous alternatives} summarizes the behavior under contiguous alternatives. A limit result under a large class of alternatives is derived in Section \ref{Section consistency}. In Section \ref{Section bootstrap procedure} we explain the implementation of the method via a parametric bootstrap procedure and prove consistency of this bootstrap-based test. We finalize the article with a Monte Carlo power study in Section \ref{Section monte carlo}, an application to observed data examples in Section \ref{Section practical applications}, and we draw conclusions and indicate open questions in Section \ref{Section conclusions}.

\section{Estimation of the parameters and asymptotic representations}
\label{Section estimators}
In this section we consider two suitable estimation methods which satisfy the requirement of scale equivariance namely the maximum likelihood (ML) and the moment (MO) estimators. For details about the estimation procedures, we refer to \cite{JKB:1995}, Chapter 15, and \cite{S:1993}, Chapter 6.

To account for the bootstrap procedure used to obtain critical values, we later on study the asymptotic behavior of $T_{n}$ under a triangular array $X_{n,1}, \dots, X_{n,n}$ of rowwise iid. random variables, where
\begin{equation*}
	X_{n,1} \sim \mbox{IG}(1, \varphi_n)
\end{equation*}
for a sequence of positive numbers $(\varphi_n)_{n \, \in \, \N}$ with $\lim_{n \, \rightarrow \, \infty} \varphi_n = \varphi > 0$. Notice that in the following we assume without loss of generality, and with respect to the scale invariance of the test statistic, that $\mu = 1$. We write $o_{\mathbb{P}}(1)$ and $O_{\mathbb{P}}(1)$ for (real-valued) random variables that converge to $0$ in probability or that are bounded in probability, respectively. For both methods we need expansions of the form
\begin{equation} \label{expansions estimators, general form}
	\sqrt{n}(\widehat{\mu}_n - 1) = \frac{1}{\sqrt{n}} \sum_{j = 1}^n \Psi_1(X_{n,j}) + \varepsilon_{n, 1}, \quad
	\sqrt{n}(\widehat{\lambda}_n - \varphi_n) = \frac{1}{\sqrt{n}} \sum_{j = 1}^n \Psi_2(X_{n,j}, \varphi_n) + \varepsilon_{n,2},
\end{equation}
where $\varepsilon_{n,j} = o_{\mathbb{P}}(1)$, $j = 1, 2$, and $\Psi_j$ are measurable functions such that the random variables $\Psi_1(X_{n,1})$ and $\Psi_2(X_{n,1}, \varphi_n)$ are centered with existing second moment, and
\begin{equation*}
	\lim_{n \, \rightarrow \, \infty} \E \Big[ \big( \Psi_1 (X_{n,1}) \big)^2 \Big] = \E \Big[ \big( \Psi_1 (X) \big)^2 \Big], \quad
	\lim_{n \, \rightarrow \, \infty} \E \Big[ \big( \Psi_2 (X_{n,1}, \varphi_n) \big)^2 \Big] = \E \Big[ \big( \Psi_2 (X, \varphi) \big)^2 \Big],
\end{equation*}
where $X \sim \mbox{IG}(1, \varphi)$.

\begin{enumerate}
	\item \textit{Maximum likelihood estimators:} Standard calculations show that
	\begin{equation*}
		\widehat{\mu}_n = \frac{1}{n} \sum_{j = 1}^{n} X_{n, j} = \overline{X}_n \quad \text{and} \quad \widehat{\lambda}_n = \left( \frac{1}{n} \sum_{j = 1}^{n} \left( \frac{1}{X_{n,j}} - \frac{1}{\overline{X}_n} \right) \right)^{-1}
	\end{equation*}
	are the ML estimators of $\mu$ and $\lambda$. The asymptotic expansions as in (\ref{expansions estimators, general form}) are derived as
	\begin{equation*}
		\sqrt{n}\big(\widehat{\mu}_n - 1\big) = \frac{1}{\sqrt{n}} \sum_{j = 1}^{n} (X_{n,j} - 1)
	\end{equation*}
	and
	\begin{equation} \label{asyexpML}
		\sqrt{n} \big(\widehat{\lambda}_n - \varphi_n\big) = \frac{1}{\sqrt{n}} \sum_{j = 1}^{n} \left( - \frac{\varphi_n^2}{X_{n,j}} - \varphi^2_n X_{n, j} + 2 \varphi_n^2 + \varphi_n \right) + o_{\mathbb{P}}(1).
	\end{equation}
	
	\item \textit{Moment estimators:} The moment estimators based on the first two moments of the inverse Gaussian distribution are
	\begin{equation*}
		\widehat{\mu}_n = \overline{X}_n \quad \text{and} \quad \widehat{\lambda}_n = \frac{\widehat{\mu}_n^3}{\frac{1}{n} \sum_{j = 1}^n X_{n, j}^2 - \widehat{\mu}_n^2},
	\end{equation*}
	with the same asymptotic expansion for $\widehat{\mu}_n$ as in the case of ML estimation, and
	\begin{equation} \label{asyexpMOM}
		\sqrt{n} (\widehat{\lambda}_n - \varphi_n) = \frac{1}{\sqrt{n}} \sum_{j = 1}^n \Big( - \varphi_n^2 X_{n, j}^2 + \left( 3 \varphi_n + 2 \varphi_n^2 \right) X_{n,j} - \varphi_n (2 + \varphi_n) \Big) + o_{\mathbb{P}}(1).
	\end{equation}
\end{enumerate}

\begin{remark} \label{Remark Estimation under alternatives}
	Note that if $X, X_1, X_2, \dotso$ are any iid. positive random variables such that $\E \big[ X + X^{-1} \big] < \infty$ (and $\E X^2 < \infty$ in case of the moment estimators), the definitions and asymptotic expansions above remain essentially the same, only $X_{n, j}$ is replaced by $X_j$, and $\varphi_n$ by $\varphi$ (with defined $\varphi$ as below). To see that the asymptotic expansions continue to hold, notice that by the scale invariance of our test statistic, we may assume that $\E X = 1$, hence $\widehat{\mu}_n \to 1$ $\mathbb{P}$-almost surely (a.s.) and $\Psi_1(X_j) = X_j - 1$ is centered. Similarly, using that
	\begin{align*}
		\widehat{\lambda}_n
		= \left( \frac{1}{n} \sum_{j = 1}^{n} \left( \frac{1}{X_{n,j}} - \frac{1}{\overline{X}_n} \right) \right)^{-1}
		\longrightarrow \left( \E X^{-1} - 1 \right)^{-1}
		=: \varphi
	\end{align*}
	or
	\begin{align*}
		\widehat{\lambda}_n
		= \frac{\widehat{\mu}_n^3}{\frac{1}{n} \sum_{j = 1}^n X_{n, j}^2 - \widehat{\mu}_n^2}
		\longrightarrow \left( \E X^{2} - 1 \right)^{-1}
		=: \varphi
	\end{align*}
	$\mathbb{P}$-a.s., as $n \to \infty$, for the ML or MO estimator, respectively, the expansions (\ref{asyexpML}) and (\ref{asyexpMOM}) are seen to remain valid with the mentioned replacements.
	
	For the ML estimators also note that since the $\mathcal{IG}$ family is a 2-parameter general exponential family, the statistic $\left(\overline{X}_n, \sum_{j = 1}^{n} \big(X_j^{-1} - \overline{X}_n^{-1}\big)\right)$ is minimal sufficient and complete. An application of the Lehmann-Scheff\'{e} theorem shows that $\widehat{\mu}_n$ and $\frac{n - 1}{n} \widehat{\lambda}_n$ are uniformly minimum variance unbiased estimators, see \cite{S:1993}. Note that $\frac{n - 1}{n} \widehat{\lambda}_n$ is the same estimator as the one considered by \cite{BS:2018}.
	
	In principle, any other estimation method which gives scale equivariant estimators that allow asymptotic expansion as above can be considered as well.
\end{remark}

\section{The limit null distribution}
\label{Section asymH0}
A suitable setting to derive asymptotic results for the test is $L^2 = L^2\big( (0, \infty), \mathcal{B}(0, \infty), w(t) \, \mathrm{d}t\big)$, the Hilbert space of Borel-measurable functions $g : (0, \infty) \rightarrow (0, \infty)$ satisfying $\int_0^\infty \big( g(t) \big)^2 \, w(t) \, \mathrm{d}t < \infty$. The inner product and norm in $L^2$ are denoted by
\begin{equation*}
	\langle g_1, g_2 \rangle_{L^2} = \int_0^\infty g_1(t) \, g_2(t) \, w(t) \, \mathrm{d}t \quad \text{and} \quad \| g \|_{L^2} = \left( \int_0^\infty \big( g(t) \big)^2 \, w(t) \, \mathrm{d}t \right)^{1/2},
\end{equation*}
for $g, g_1, g_2 \in L^2$. We assume the triangular array $X_{n,1}, \ldots, X_{n,n}$ from Section \ref{Section estimators}. In particular, recall that\linebreak $X_{n, 1} \sim \mbox{IG}(1, \varphi_n)$ with $\varphi_n \to \varphi > 0$, and define
\begin{equation*}
	V_n(t) = \frac{1}{\sqrt{n}} \sum\limits_{j = 1}^{n} \left[ \frac{1}{2} \left( \frac{\widehat{\lambda}_n}{\widehat{\mu}_n^2} + \frac{3}{X_{n, j}} - \frac{\widehat{\lambda}_n}{X_{n, j}^2} \right) \min\{ X_{n, j}, t \} - \mathds{1}\{X_{n, j}\leq t\} \right], \quad t > 0,
\end{equation*}
for one of the two types of estimators $\widehat{\mu}_n$ and $\widehat{\lambda}_n$ as in Section \ref{Section estimators}. Note that $V_n(\cdot)$ is a random element of $L^2$, and after a simple change of variable we have
\begin{equation*}
	T_{n} = \frac{1}{\widehat{\mu}_n} \int_0^\infty \big( V_n(t) \big)^2 \, w \left( \frac{t}{\widehat{\mu}_n} \right) \, \mathrm{d}t.
\end{equation*}
The first step of the proof of Theorem 3 by \cite{BE:2019:2} yields the following result [using assumption (\ref{weightcond})].
\begin{lemma} \label{Lemma approx of rescaled weight}
	Assume that $\| V_n \|_{L^2}^2$, $n \in \N$, is a tight sequence. Then $T_n = \frac{1}{\widehat{\mu}_n} \, \| V_n \|_{L^2}^2 + o_{\mathbb{P}}(1)$.
\end{lemma}
Now define
\begin{align*}
	V^*_n(t)
	&= \frac{1}{\sqrt{n}} \sum\limits_{j = 1}^{n} \left[ \frac{1}{2} \left( \varphi_n + \frac{3}{X_{n, j}} - \frac{\varphi_n}{X_{n, j}^2} \right) \min\{ X_{n, j}, t \} - \mathds{1}\{X_{n, j}\leq t\} \right] \\
	&~~~ + \frac{\sqrt{n}}{2} \left( \frac{\widehat{\lambda}_n}{\widehat{\mu}_n^2} - \varphi_n \right) \, \E \big[ \min\{ X, t \} \big] - \frac{\sqrt{n}}{2} \big( \widehat{\lambda}_n - \varphi_n \big) \, \E \left[ \frac{\min\{ X, t \}}{X^2} \right] , \quad t > 0.
\end{align*}
Tedious but straightforward approximations yield the following relation.
\begin{lemma} \label{Lemma approx of integrand of the test statistic}
	As $n \to \infty$, we have $\| V_n - V_n^* \|_{L^2} \stackrel{\mathbb{P}}{\longrightarrow} 0$.
\end{lemma}
As these calculations provide no essential insights, we omit them here and refer to \cite{BE:2019:2} where similar arguments are stated to prove Theorem 3 in their work. It is an easy corollary of Lemma \ref{Lemma approx of rescaled weight} that the test statistic can be written as follows.
\begin{corollary} \label{Corollary approx of the test statistic via approx integrand}
	Assume that $V_n^*$, $n \in \N$, is a tight sequence in $L^2$. Then $T_n = \| V_n^* \|_{L^2}^2 + o_{\mathbb{P}}(1)$.
\end{corollary}
Next, we write $V_n^*$ as a sum of iid. random elements of $L^2$, and use a central limit theorem for triangular arrays to conclude that $V_n^*$ is a tight sequence in $L^2$ (thus rendering Corollary \ref{Corollary approx of the test statistic via approx integrand} applicable) and derive the limit distribution. To this end, notice that we obtain from (\ref{expansions estimators, general form})
\begin{align*}
	\sqrt{n} \left( \frac{\widehat{\lambda}_n}{\widehat{\mu}_n^2} - \varphi_n \right)
	= - 2 \, \varphi_n \frac{1}{\sqrt{n}} \sum_{j = 1}^{n} \Psi_1(X_{n, j}) + \frac{1}{\sqrt{n}} \sum_{j = 1}^{n} \Psi_2(X_{n, j}, \varphi_n) + o_{\mathbb{P}}(1) ,
\end{align*}
and thus, defining
\begin{align*}
	V^{**}_n(t)
	&= \frac{1}{\sqrt{n}} \sum\limits_{j = 1}^{n} W_{n, j}(t) ,
\end{align*}
where
\begin{align*}
	W_{n, j}(t)
	&= \frac{1}{2} \left( \varphi_n + \frac{3}{X_{n, j}} - \frac{\varphi_n}{X_{n, j}^2} \right) \min\{ X_{n, j}, t \} - \mathds{1}\{X_{n, j}\leq t\} \\
	&~~~~~~~~ - \varphi_n \Psi_1(X_{n, j}) \, \E \big[ \min\{ X, t \} \big] + \frac{1}{2} \, \Psi_2(X_{n, j}, \varphi_n) \, \E \Big[ \min\{ X, t \} \big( 1 - X^{-2} \big) \Big],
\end{align*}
we get $\| V_n^* - V_n^{**} \|_{L^2} = o_{\mathbb{P}}(1)$. Notice that $W_{n, 1}, \dots, W_{n, n}$ are iid. random elements of $L^2$ with $\E W_{n, 1} = 0$ (by Theorem \ref{Theorem characterization}) and $\E \| W_{n, j} \|_{L^2}^2 < \infty$ (using the assumptions on $\Psi_1$, $\Psi_2$, and $w$). This asymptotic representation leads to the following theorem which gives the limit null distribution of $T_n$.
\begin{theorem} \label{Theorem asyH0}
	Under the standing assumptions there exists a centered Gaussian element, $\mathcal{W}$, of $L^2$ with covariance kernel
	\begin{align*}
		\mathcal{K}_\varphi (s,t) = \E \Big[ g_\varphi(s, X) \, g_\varphi(t, X) \Big], \quad s, t > 0,
	\end{align*}
	where
	\begin{align*}
		g_\varphi(s, x) &= \frac{1}{2} \left( \varphi + \frac{3}{x} - \frac{\varphi}{x^2} \right) \min\{ x, s \} - \mathds{1}\{x \leq s\} - \varphi \Psi_1(x) \, \E \big[ \min\{ X, s \} \big] \\
		&~~~+ \frac{1}{2} \Psi_2(x, \varphi) \, \E \big[ \min\{ X, s \} \big( 1 - X^{-2} \big) \big], \quad x, s > 0,
	\end{align*}
	such that
	\begin{equation*}
		T_{n} \stackrel{\mathcal{D}}{\longrightarrow}
		\left\|\mathcal{W}\right\|_{L^2}^2 , \quad \text{as } n \to \infty,
	\end{equation*}
	where $\stackrel{\mathcal{D}}{\longrightarrow}$ denotes convergence in distribution.
\end{theorem}
\noindent
\begin{prf}
	We verify condition (a) from Lemma 3.1 and condition (ii) from Remark 3.3 of \cite{CW:1998} to apply the corresponding central limit theorem for triangular arrays in Hilbert spaces to the array $\left\{ W_{n, j} \, \big| \, j \in \{ 1, \dots, n \} ,\linebreak \, n \in \N \right\}$. As argued above, these random elements are centered and have finite second moment. Moreover, the limit
	\begin{align} \label{limit results for the CLT}
		\lim_{n \, \rightarrow \, \infty} \E \left\| \frac{1}{\sqrt{n}} \sum_{j = 1}^{n} W_{n, j} \right\|_{L^2}^2
		= \lim_{n \, \rightarrow \, \infty} \int_0^\infty \E\Big[ \big( W_{n, 1}(s) \big)^2 \Big] w(s) \, \mathrm{d}s
	\end{align}
	exists and is finite. The boundedness is seen by checking that the '$\limsup$' of the sequence is finite, using the existence of moments of inverse Gaussian random variables, the assumptions on the weight function, and the asymptotic expansions of the estimators, as well as H\"{o}lder's inequality. That very calculation also justifies the exchange of the limit '$\lim_{n \, \rightarrow \, \infty}$' with the first integral. Then to see that the limit in (\ref{limit results for the CLT}) exists, the convergence of the term $\lim_{n \, \rightarrow \, \infty} \E \left[ \big( W_{n, 1}(s) \big)^2 \right]$ is shown by proving that the separate terms in the expectation (after carrying out the square) are uniformly integrable, utilizing the explicit form of $\Psi_1$, $\Psi_2$ for the ML or MO estimators, and the fact that $\sup_{n \, \in \, \N} \E X_{n, 1}^8 < \infty$, $\sup_{n \, \in \, \N} \E X_{n, 1}^{-4} < \infty$. The classical Lindeberg central limit theorem for arrays implies
	\begin{align*}
		\frac{1}{\sqrt{n}} \sum_{j = 1}^{n} \big\langle W_{n, j}, g \big\rangle_{L^2}
		\stackrel{\mathcal{D}}{\longrightarrow} \mathcal{N}\big( 0, \sigma_\varphi^2(g) \big)
	\end{align*}
	for any $g \in L^2 \setminus \{ 0 \}$, with $\sigma_\varphi^2(g) = \lim_{n \, \rightarrow \, \infty} \E\big[ \langle W_{n, 1}, g \rangle_{L^2}^2 \big]$ which is finite by the finiteness of the term in (\ref{limit results for the CLT}). From the limit theorem by \cite{CW:1998} we obtain the convergence $V_n^{**} = n^{- 1/2} \sum_{j = 1}^{n} W_{n, j} \stackrel{\mathcal{D}}{\longrightarrow} \mathcal{W}$, where $\mathcal{W}$ is as in the statement of the theorem, and where we used
	\begin{align*}
		\sigma_\varphi^2(g)
		= \int_0^\infty \int_0^\infty \mathcal{K}_\varphi(s, t) \, g(s) \, g(t) \, w(s) \, w(t) \, \mathrm{d}s \, \mathrm{d}t.
	\end{align*}
	Since $\| V_n^* - V_n^{**} \|_{L^2} = o_{\mathbb{P}}(1)$, $\{ V_n^* \}_{n \, \in \, \N}$ is a tight sequence in $L^2$, and Corollary \ref{Corollary approx of the test statistic via approx integrand} implies the claim.
\end{prf}

\section{Contiguous alternatives}
\label{Section contiguous alternatives}
In this section we summarize the asymptotic behavior under contiguous alternatives to a fixed representative of $\mathcal{IG}$. As the proof of the subsequent results is almost identical to the arguments used by \cite{BE:2019:2} in Theorem 4, we omit the details and only state the results. Let $f_0$ denote the density of the $\mbox{IG}(1, \varphi)$ distribution for a fixed $\varphi > 0$, and let $c : (0, \infty) \to \R$ be a measurable, bounded function satisfying $\int_0^\infty c(x) \, f_0(x) \, \mathrm{d}x = 0$. Further, let $Z_{n, 1}, \dots, Z_{n, n}$, $n \in \N$, be a triangular array of rowwise iid. random variables with Lebesgue density
\begin{equation*}
	f_n(x) = f_0(x) \left( 1 + \frac{c(x)}{\sqrt{n}} \right), \quad x > 0,
\end{equation*}
where we assume $n$ to be large enough to ensure the non-negativity of $f_n$. The $n$-fold product measure of $f_n(x) \mathrm{d}x$ is contiguous to the $n$-fold product measure of $f_0(x) \mathrm{d}x$ in terms of \cite{C:1960}.
\begin{theorem}\label{Theorem contiguous alternatives}
	Under the stated assumptions we have
	\begin{equation*}
		T_{n} = T_{n}(Z_{n, 1}, \dots, Z_{n, n}) \stackrel{\mathcal{D}}{\longrightarrow} \big\| \mathcal{W} + \zeta \big\|_{L^2}^2 , \quad \text{as } n \rightarrow \infty,
	\end{equation*}
	where $\zeta(\cdot) = \int_0^\infty g_\varphi(\cdot, x) \, c(x) \, f_0(x) \, \mathrm{d}x$, and $\mathcal{W}$ is the centered Gaussian element of $L^2$ from Theorem \ref{Theorem asyH0} with the function $g_\varphi(\cdot, \cdot)$ related to the covariance kernel of $\mathcal{W}$ as in the previous section.
\end{theorem}

We thus conclude that our test has non-trivial power against contiguous alternatives of the considered kind.

\section{A functional law of large numbers}
\label{Section consistency}
In the next section we explain how the test is implemented using a parametric bootstrap procedure. We also show that Theorem \ref{Theorem asyH0}, that is, the result on the limit null distribution from Section \ref{Section asymH0}, can be used to prove that the bootstrap is asymptotically valid. However, we would also like to show that the bootstrap-based test is consistent against fixed alternatives. To achieve this goal, we provide another limit result, this time considering the setting under alternative distributions. Let $X$ be any positive random variable with $\E X < \infty$ and $\E X^{-1} <\infty$ (and $\E X^2 < \infty$ if moment estimators are used), and let $X_1, X_2, \dotso$ be iid. copies of $X$. We assume that either the ML or the moment estimators are considered, and we thus have, from Remark \ref{Remark Estimation under alternatives}, that
\begin{equation*}
	\big(\widehat{\mu}_n, \widehat{\lambda}_n\big) \longrightarrow (\mu, \lambda)
\end{equation*}
$\mathbb{P}$-a.s., as $n \to \infty$, for some $\mu, \lambda > 0$ (where we can specify $\mu = \E X$). In view of the scale invariance of $T_{n}$, we assume $\mu = \E X = 1$, and have $\widehat{\varphi}_n = \widehat{\lambda}_n / \widehat{\mu}_n \longrightarrow \lambda / \mu = \lambda =: \varphi$ $\mathbb{P}$-a.s., as $n \to \infty$, where $\varphi$ may be specified as in Remark \ref{Remark Estimation under alternatives}.
\begin{theorem}\label{Theorem consistency}
	As $n \rightarrow \infty$, we have
	\begin{equation*}
		\frac{T_{n}}{n}
		\stackrel{\mathbb{P}}{\longrightarrow} \Delta_{\varphi}
		= \int_0^{\infty} \left| \E \left[ \frac{1}{2} \left( \varphi + \frac{3}{X} - \frac{\varphi}{X^2} \right) \min\{ X, t \} \right] - \mathbb{P}(X \leq t) \right|^2 w(t) \, \mathrm{d}t.
	\end{equation*}
\end{theorem}
\noindent
\begin{prf}
	Define
	\begin{align*}
		\widetilde{V}_n(t)
		= \frac{1}{n} \sum_{j = 1}^{n} \left[ \frac{1}{2} \left( \frac{\widehat{\lambda}_n}{\widehat{\mu}_n^2} + \frac{3}{X_j} - \frac{\widehat{\lambda}_n}{X_j^2} \right) \min \{ X_j, t \} - \mathds{1}\{ X_j \leq t \} \right] , \quad t > 0.
	\end{align*}
	With multiple use of the triangle inequality, $\left\{ \big\| \widetilde{V}_n \big\|_{L^2}^2 \right\}_{n \, \in \, \N}$ is easily seen to be a tight sequence of random variables. A statement identical to Lemma \ref{Lemma approx of rescaled weight} yields
	\begin{align*}
		\frac{T_n}{n}
		= \frac{1}{\widehat{\mu}_n} \int_0^\infty \big( \widetilde{V}_n(t) \big)^2 \, w\left( \frac{t}{\widehat{\mu}_n} \right) \mathrm{d}t
		= \big\| \widetilde{V}_n \big\|_{L^2}^2 + o_{\mathbb{P}}(1).
	\end{align*}
	Upon setting
	\begin{align*}
		\widetilde{V}_{lim}^{(\varphi)}(t)
		= \E \left[ \frac{1}{2} \left( \varphi + \frac{3}{X} - \frac{\varphi}{X^2} \right) \min\{ X, t \} \right] - \mathbb{P}(X \leq t), \quad t > 0,
	\end{align*}
	we have $\Delta_{\varphi} = \big\| \widetilde{V}_{lim}^{(\varphi)} \big\|_{L^2}^2$, and therefore
	\begin{align*}
		\left| \frac{T_n}{n} - \Delta_{\varphi} \right|
		&= \Big| \big\| \widetilde{V}_n \big\|_{L^2}^2 - \big\| \widetilde{V}_{lim}^{(\varphi)} \big\|_{L^2}^2 \Big| + o_{\mathbb{P}}(1) \\
		&\leq \Big( \big\| \widetilde{V}_n \big\|_{L^2} + \sqrt{\Delta_\varphi} \Big) \, \big\| \widetilde{V}_n - \widetilde{V}_{lim}^{(\varphi)} \big\|_{L^2} + o_{\mathbb{P}}(1).
	\end{align*}
	The first term in the product on the right hand side of this inequality is tight, while the second term is seen to converge to $0$ almost surely by an easy calculation and standard Glivenko-Cantelli arguments (see \cite{BE:2019:2}, Theorem 5, for some further insight).
\end{prf}

Because of Theorem \ref{Theorem characterization} the limit in Theorem \ref{Theorem consistency} is $0$ if, and only if, $X$ follows some inverse Gaussian distribution $\mbox{IG}(1, \varphi)$, and it is strictly greater than $0$ otherwise.

\section{The bootstrap procedure and a consistency result}
\label{Section bootstrap procedure}
In practice, to carry out the test at some level $\alpha \in (0, 1)$, we suggest a parametric bootstrap procedure, as the distribution of the test statistic depends on an unknown parameter and is very complicated. The approach is as follows. Given a sample $X_1, \dots, X_n$ of iid. positive random variables with $\E \big[ X_1 + X_1^{-1} \big] < \infty$ (and $\E X_1^2 < \infty$ when moment estimation is used), calculate $T_n(X_1, \dots, X_n)$, using for instance the explicit formulae from Section \ref{Subsection computational form}. Also calculate the estimators $\widehat{\mu}_n(X_1, \dots, X_n)$ and $\widehat{\lambda}_n(X_1, \dots, X_n)$, and put $\widehat{\varphi}_n(X_1, \dots, X_n) = \widehat{\lambda}_n(X_1, \dots, X_n) / \widehat{\mu}_n(X_1, \dots, X_n)$. Conditional on this value of $\widehat{\varphi}_n$, generate $b$ samples of size $n$ from the $\mbox{IG}(1, \widehat{\varphi}_n)$-distribution, and calculate the test statistic for each of them, with the parameters estimated from the bootstrap sample in every instance. This yields the values $T_{n, 1}^{*}, \dots, T_{n, b}^{*}$. Define the empirical distribution function
\begin{align*}
	F_{n, b}^{*}(t) = \frac{1}{b} \sum_{k = 1}^{b} \mathds{1}\big\{ T_{n, k}^* \leq t \big\}, \quad t > 0,
\end{align*}
and the empirical $(1 - \alpha)$-quantile, $c_{n, b}^{*}(\alpha) = F_{n, b}^{* -1}(1 - \alpha)$, of $T_{n, 1}^{*}, \dots, T_{n, b}^{*}$. The hypothesis (\ref{H0}) is rejected if $T_n(X_1, \dots, X_n) > c_{n, b}^{*}(\alpha)$. Denote by $F_\varphi$ the distribution function of $\| \mathcal{W} \|_{L^2}^2$ under the $\mbox{IG}(1, \varphi)$-distribution as in Theorem \ref{Theorem asyH0}, where $\varphi$ is the (almost sure) limit of $\widehat{\varphi}_n$ which exists by Remark \ref{Remark Estimation under alternatives}. It is straightforward to adapt the methods from \cite{Hen:1996}, applying Theorem \ref{Theorem asyH0}, to prove that
\begin{align*}
	\sup_{t \, > \, 0} \big| F_{n, b}^{*}(t) - F_\varphi(t) \big|
\stackrel{\mathbb{P}}{\longrightarrow} 0, \quad \text{as } n, b \to \infty,
\end{align*}
and thus $c_{n, b}^{*}(\alpha) \stackrel{\mathbb{P}}{\longrightarrow} F_\varphi^{-1}(1 - \alpha) \in (0, \infty)$, as $n, b \to \infty$. Hence, if $X_1 \sim \mbox{IG}(\mu, \lambda)$ for some $\mu, \lambda > 0$, then
\begin{align*}
	\mathbb{P}\Big( T_n(X_1, \dots, X_n) > c_{n, b}^{*}(\alpha) \Big) \longrightarrow \alpha, \quad \text{as } n, b \to \infty.
\end{align*}
We conclude that a given level of significance is attained in the limit, as both the sample size and the chosen bootstrap size approach infinity. The bootstrap procedure is thus asymptotically valid. \\

Now suppose that $X_1$ from above does not follow any inverse Gaussian law. Then the limit $\Delta_\varphi$ figuring in Theorem \ref{Theorem consistency} is strictly positive. Consequently, Theorem \ref{Theorem consistency} and the results on the bootstrap critical values above imply
\begin{align*}
	\mathbb{P}\Big( T_n(X_1, \dots, X_n) > c_{n, b}^{*}(\alpha) \Big)
	\longrightarrow 1, \quad \text{as } n, b \to \infty,
\end{align*}
that is, our test is consistent (in the bootstrap setting) against any such alternative distribution. We suggest using the above bootstrap procedure when the test is applied in practice. In the extensive power approximation in the following section, the procedure becomes very demanding due to the high number of Monte Carlo runs. To accelerate the computations, we employ in that simulation study the so-called warp-speed bootstrap, see \cite{GPW:2013}, as is explained in Section \ref{Section monte carlo}.

\section{Monte Carlo study}
\label{Section monte carlo}

This section compares the finite sample power performance of the newly proposed test to that of competing tests for the inverse Gaussian distribution. Below, we consider the computational form of our test statistic. Then various existing goodness-of-fit tests for the inverse Gaussian distribution are discussed. Finally, the power calculations, including the considered alternative distributions and the warp-speed bootstrap methodology, are detailed. Sample sizes of $30$ and $50$ are considered throughout.

\subsection{Computational form}
\label{Subsection computational form}
We start with computationally stable representations of the test statistic for the weight functions $w_a(t) = \exp(- a t)$ and $\widetilde{w}_a(t) = \exp(- a t^2)$, $t > 0$. For $w_a$ we define
\begin{eqnarray*}
	h_{1,a}(s, t) &=& \left\{ \begin{array}{cc} \frac{1}{a^3} \left( 2 - e^{-as} \left( (a s + 1)^2 + 1 \right) \right) + \frac{s}{a^2} \left( e^{- a s} (a s + 1) - e^{- a t} (a t + 1) \right) + \frac{s t}{a} e^{- a t}, & s \leq t, \\[1mm] \frac{1}{a^3} \left( 2 - e^{- a t} \left( (a t + 1)^2 + 1 \right) \right) + \frac{t}{a^2} \left( e^{- a t} (a t + 1) - e^{- a s} (a s + 1) \right) + \frac{s t}{a} e^{- a s}, & s > t, \end{array} \right. \\
	h_{2,a}(s, t) &=& \left\{ \begin{array}{cc} \frac{s}{a} e^{ - a t}, & s \leq t, \\[1mm] \frac{1}{a^2} \left( e^{- a t} (a t + 1) - e^{- a s} (a s + 1) \right) + \frac{s}{a} e^{- a s}, & s > t, \end{array} \right. \\
\end{eqnarray*}
and get a numerically stable version of the test statistic,
\begin{eqnarray*}
	T_{n, a} &=& \frac{1}{4n} \sum_{j, k = 1}^{n} \left( \widehat{\varphi}_n + \frac{3}{Y_{n, j}} - \frac{\widehat{\varphi}_n}{Y_{n, j}^2} \right) \left( \widehat{\varphi}_n + \frac{3}{Y_{n, k}} - \frac{\widehat{\varphi}_n}{Y_{n, k}^2} \right) h_{1, a}(Y_{n, j}, Y_{n, k})\\ &&
	~~~~~~~~~~~~-2 \left( \widehat{\varphi}_n + \frac{3}{Y_{n, j}} - \frac{\widehat{\varphi}_n}{Y_{n, j}^2} \right) h_{2, a}(Y_{n, j}, Y_{n, k}) \\ &&
	~~~~~~~~~~~~-2 \left( \widehat{\varphi}_n + \frac{3}{Y_{n, k}} - \frac{\widehat{\varphi}_n}{Y_{n, k}^2} \right) h_{2, a}(Y_{n, k}, Y_{n, j}) + \frac{4}{a} e^{- a \max(Y_{n, j}, Y_{n, k})},
\end{eqnarray*}
where $Y_{n, 1}, \dots, Y_{n, n}$ and $\widehat{\varphi}_n$ are as in Section \ref{Section introduction}. For the second weight, $\widetilde{w}_a$, define
\begin{eqnarray*}
	\widetilde{h}_{1, a}(s, t) &=& \left\{ \begin{array}{cc} \frac{\sqrt{\pi / a^3}}{4} - \frac{a}{2} \sqrt{\tfrac{\pi}{a^5}} \, \Phi(- \sqrt{2 a} s) - \frac{s}{2 a} e^{- a t^2} + s t \sqrt{\tfrac{\pi}{a}} \, \Phi(- \sqrt{2 a} t), & s \leq t, \\
	\frac{\sqrt{\pi / a^3}}{4} - \frac{a}{2} \sqrt{\tfrac{\pi}{a^5}} \, \Phi(- \sqrt{2 a} t) - \frac{t}{2 a} e^{- a s^2} + s t \sqrt{\tfrac{\pi}{a}} \, \Phi(- \sqrt{2 a} s), & s > t, \end{array} \right. \\
	\widetilde{h}_{2, a}(s, t) &=& \left\{ \begin{array}{cc} s \, \sqrt{\tfrac{\pi}{a}} \, \Phi(- \sqrt{2 a} t), & s \leq t, \\
	\frac{1}{2 a} \left( e^{- a t^2} - e^{- a s^2} \right) + s \, \sqrt{\tfrac{\pi}{a}} \, \Phi(- \sqrt{2 a} s), & s > t, \end{array} \right. \\
\end{eqnarray*}
where $\Phi$ denotes the distribution function of the standard normal distribution. Then we have a numerically stable version of the corresponding test statistic, namely
\begin{eqnarray*}
	\widetilde{T}_{n, a} &=& \frac  {1}{4 n} \sum_{j, k = 1}^{n} \left( \widehat{\varphi}_n + \frac{3}{Y_{n, j}} - \frac{\widehat{\varphi}_n}{Y_{n, j}^2} \right) \left( \widehat{\varphi}_n + \frac{3}{Y_{n, k}} - \frac{\widehat{\varphi}_n}{Y_{n, k}^2} \right) \widetilde{h}_{1, a}(Y_{n, j}, Y_{n, k}) \\ &&
	~~~~~~~~~~~~-2 \left( \widehat{\varphi}_n + \frac{3}{Y_{n, j}} - \frac{\widehat{\varphi}_n}{Y_{n, j}^2} \right) \widetilde{h}_{2, a}(Y_{n, j}, Y_{n, k}) \\ &&
	~~~~~~~~~~~~-2 \left( \widehat{\varphi}_n + \frac{3}{Y_{n, k}} - \frac{\widehat{\varphi}_n}{Y_{n, k}^2} \right) \widetilde{h}_{2, a}(Y_{n, k}, Y_{n, j}) + 4 \sqrt{\frac{\pi}{a}} \, \Phi\left( - \sqrt{2 a} \max(Y_{n, j}, Y_{n, k}) \right).
\end{eqnarray*}

\subsection{Existing tests of fit for the inverse Gaussian distribution}
\label{Subsection competing tests}

The class of competing tests we consider comprises several classical tests as well as more recent tests. We choose the following procedures:

\begin{enumerate}
	\item The Kolmogorov-Smirnov test,

	\item the Cram\'{e}r-von Mises test,

	\item the Anderson-Darling test,

	\item two tests proposed by \cite{HK:2002},

	\item a recent test by \cite{VG:2015}.
\end{enumerate}

Below, we briefly provide the details of these tests. The first three tests are well-known and we only provide the computational form of the test statistic in each case. The remaining three tests are considered in more detail. The test by \cite{VG:2015} is very recent, while the tests by \cite{HK:2002} are included due to their impressive power performance in previous empirical studies. For a recent literature overview concerning the existing tests for the inverse Gaussian distribution, see \cite{KK:2012}.

\subsubsection{Classical tests}
\label{Subsection classical tests}

Let $X_{\left( j \right)}$ denote the $j^{\text{th}}$ order statistic of $X_{1}, \dots, X_{n}$, and let $\widehat{F} \left( x \right) = F \big( x; \, \widehat{\mu}_{n}, \widehat{\lambda}_{n} \big) $, where $F$ is the distribution function of the inverse Gaussian distribution. For each of the following tests, the null hypothesis is rejected for large values of the test statistic.

\begin{enumerate}
	\item \textit{Kolmogorov-Smirnov test:} The form of the test statistic is
	\begin{equation*}
		KS = \max \left( D^{+}, D^{-} \right) ,
	\end{equation*}%
	where $D^{+} = \max_{j \, = \, 1, \dots, n} \left( \frac{j}{n} - \widehat{F} \left( X_{\left( j \right)} \right) \right)$ and $D^{-} = \max_{j \, = \, 1, \dots, n} \left( \widehat{F} \left( X_{\left( j \right)} \right) - \frac{j-1}{n} \right)$.

	\item \textit{Cram\'{e}r-von Mises test:} In this case, the test statistic is
	\begin{equation*}
		CM = \frac{1}{12n} + \sum_{j = 1}^{n} \left( \widehat{F} \left( X_{\left( j \right)} \right) - \frac{2j - 1}{2n} \right)^{2}.
	\end{equation*}

	\item \textit{Anderson-Darling test:} The computational form of the Anderson-Darling test statistic is
	\begin{equation*}
		AD = - n - \frac{1}{n} \sum_{j = 1}^{n} \left[ \left( 2j - 1 \right) \log \widehat{F} \left( X_{\left( j \right)} \right) + \big( 2 \left( n - j \right) + 1 \big) \log \left( 1 - \widehat{F} \left( X_{\left( j \right)} \right) \right) \right] .
	\end{equation*}
\end{enumerate}

\subsubsection{Tests proposed by \cite{HK:2002}}
\label{Subsection test Henze Klar}

\cite{HK:2002} proposed two classes of tests for the inverse Gaussian distribution based on the empirical Laplace transform. The Laplace transform of the $\mbox{IG} \left( \mu , \lambda \right) $ distribution is given by
\begin{equation*}
	L\left( t \right) = \E \big[ \exp \left( -t X \right) \big] = \exp \left( \frac{\lambda}{\mu} \left( 1 - \sqrt{1 + \frac{2 \, \mu^{2} \, t}{\lambda}} \right) \right) , \quad t \geq 0.
\end{equation*}%
As a result, the Laplace transform of the inverse Gaussian distribution satisfies the characteristic differential equation
\begin{equation} \label{Laplace diif eq}
	\mu \, L \left( t \right) + \sqrt{1 + \frac{2 \, \mu^{2} \, t}{\lambda}} \, L^{\prime} \left( t \right) = 0, \quad t \geq 0,
\end{equation}%
subject to the initial condition $L \left( 0 \right) = 1$. The empirical Laplace transform is given by
\begin{equation*}
	L_{n} \left( t \right) = \frac{1}{n} \sum_{j = 1}^{n} \exp \left( -t X_{j} \right) .
\end{equation*}%
Under the assumption that $X_{1}, \dots, X_{n}$ are realized from an inverse Gaussian distribution, (\ref{Laplace diif eq}) suggests that
\begin{equation*}
	\widetilde{\varepsilon}_{n} \left( t \right) = \widehat{\mu}_{n} \, L_{n} \left( t \right) + \sqrt{1 + \frac{2 \, \widehat{\mu}_{n}^{2} \, t}{\widehat{\lambda}_{n}}} \, L_{n}^{\prime} \left( t \right)
\end{equation*}%
is close to zero for each value of $t$. The proposed class of test statistics thus is
\begin{equation*}
	HK^{(1)}_{n, a} = \frac{n}{\widehat{\mu}_{n}} \int_{0}^{\infty} \big( \widetilde{\varepsilon}_{n}(t) \big)^{2} \exp \left( -a \, \widehat{\mu}_{n} \, t \right) \mathrm{d}t,
\end{equation*}
where $\widehat{\mu}_{n}$ and $\widehat{\lambda}_{n}$ denote the maximum likelihood estimates of $\mu $ and $\lambda$, respectively, and $a \geq 0$ is a tuning parameter. Due to the intractability of some of the calculations required for the implementation of the test statistic, the authors recommend the use of the exponentially scaled complementary error function
\begin{equation*}
	\mathrm{erfce} \left( x \right) = \frac{2 \exp \left( x^{2} \right)}{\pi} \int_{x}^{\infty} \exp \left( - t^{2} \right) \mathrm{d}t
\end{equation*}%
rather than the distribution  function of a Gaussian random variable in the value of the test statistic. Note that $\mathrm{erfce}(x)$ tends to $\infty$ for small values of $x$. Furthermore, for sufficiently large values of $x$, the numerical evaluation of $\mathrm{erfce}(x)$ breaks down, since the value of the integral and the exponential function are calculated to be $0$ and $\infty$, respectively, by all standard statistical software packages. The latter difficulty can be overcome by noting that $\lim_{x \rightarrow \infty} \mathrm{erfce}(x) = 0$. As a result, the use of the $\mathrm{erfce}$ function reduces the numerical problems encountered when implementing the tests proposed in \cite{HK:2002} without removing these difficulties altogether.

Letting $\widehat{\varphi}_{n} = \widehat{\lambda}_{n} / \widehat{\mu}_{n}$ and $Y_{n, j} = X_{j} / \widehat{\mu}_{n}$ as in Section \ref{Section introduction}, and $\widehat{Z}_{j k} = \widehat{\varphi}_{n} \cdot \left( Y_{n, j} + Y_{n, k} + a \right) $, the test statistic can be expressed in a tractable form, namely
\begin{equation*}
	HK^{(1)}_{n, a} = \frac{\widehat{\varphi}_{n}}{n} \sum_{j, k = 1}^{n} \widehat{Z}_{j k}^{-1} \left\{ 1 - (Y_{n, j} + Y_{n, k}) \left( 1 + \sqrt{\frac{\pi}{2 \widehat{Z}_{j k}}} \, \mathrm{erfce} \left( \sqrt{\frac{\widehat{Z}_{j k}}{2}} \right) \right) + \left( 1 + \frac{2}{\widehat{Z}_{j k}}\right) Y_{n, j} \, Y_{n, k} \right\} .
\end{equation*}%
The null hypothesis is rejected for large values of $HK^{(1)}_{n, a}$. Based on the power performance of this class of tests, the authors recommend the use of $a = 0$. This recommendation is met in order to obtain the numerical results shown below.

\cite{HK:2002} also proposed a second, more immediate, class of tests based on the empirical Laplace transform. This second class of tests is based on the difference between the Laplace transform of the $\mbox{IG}\big( \widehat{\mu}_{n}, \widehat{\lambda}_{n} \big) $ distribution (denoted by $\widehat{L}$) and the empirical Laplace transform:
\begin{equation*}
	HK^{(2)}_{n, a} = n \, \widehat{\mu}_{n} \int_{0}^{\infty} \left( L_{n} \left( t \right) - \widehat{L} \left( t \right) \right)^{2} \exp \left( - a \, \widehat{\mu}_{n} \, t \right) \mathrm{d}t,
\end{equation*}%
for some $a \geq 0$. Two distinct computational forms for the test statistic are obtained, distinguishing the cases $a = 0$ and $a > 0$. Again, the recommended value of the tuning parameter is $a = 0$. In this case, the test statistic can be expressed as
\begin{equation*}
	HK^{(2)}_{n, 0} = \frac{1}{n} \sum_{j, k = 1}^{n} Z_{j k}^{-1} - 2 \sum_{j = 1}^{n} Y_{n, j}^{-1} \left\{ 1 - \sqrt{\frac{\pi \, \widehat{\varphi}_{n}}{2 Y_{n, j}}} \, \mathrm{erfce} \left( \frac{\widehat{\varphi}_{n}^{1/2} \left( Y_{n, j} + 1 \right)}{\sqrt{2 Y_{n, j}}} \right) \right\} + n \, \frac{1 + 2 \widehat{\varphi}_{n}}{4 \widehat{\varphi}_{n}},
\end{equation*}
where $Z_{j k} = Y_{n, j} + Y_{n ,k}$. The hypothesis that the data is realized from an inverse Gaussian distribution is rejected for large values of the test statistic.

\subsubsection{The test of \cite{VG:2015}}
\label{Subsection test Villasenor Gonzales-Estrada}

\cite{VG:2015} introduced three goodness-of-fit tests for the inverse Gaussian distribution, the most
powerful of which is discussed below. Using the same notation as the mentioned authors, it can be shown that if $X\sim \mbox{IG}\left( \mu , \lambda \right)$, then
\begin{equation*}
	Z^{\left( \mu \right)} = \frac{\left( X - \mu \right)^{2}}{X}
\end{equation*}%
follows a Gamma distribution with shape parameter $1/2$ and scale parameter $2 \mu^{2} / \lambda $. A goodness-of-fit test for the inverse Gaussian distribution can be constructed using the moment estimator of $Cov \big( X, Z^{\left( \mu \right)} \big)$, where $\mu$ is estimated by $\widehat{\mu}_n = \overline{X}_{n}$. The suggested test statistic can be written as
\begin{equation*}
	VG_{n} = \sqrt{\frac{n \widehat{\lambda}_{n}}{6 \widehat{\mu}_{n}}} \left( \frac{\widehat{\lambda}_{n} S_{n}^{2}}{\widehat{\mu}_{n}^{3}} - 1 \right) ,
\end{equation*}%
where $S_{n}^{2}$ denotes the sample variance of $X_{1}, \dots, X_{n}$, and where $\widehat{\lambda}_n$ is the ML estimator of $\lambda$. The asymptotic distribution of $VG_{n}$ is standard normal, and the null hypothesis is rejected for large values of $\left \vert VG_{n} \right \vert$.

\subsection{Power calculations}
\label{Subsection power calculations}

Table \ref{Table alternative distributions} shows the alternative distributions considered in the empirical study below. Each of the listed distributions is investigated for various parameter values. The powers of the tests against the inverse Gaussian distribution with mean parameter $1$ and shape parameter $\theta$, denoted in Tables \ref{Table power results n = 30} and \ref{Table power results n = 50} by $\mbox{IG} \left( \theta \right)$, are also calculated for several values of $\theta$ in order to evaluate the empirical size of all competing tests.

\begin{table}
\centering
\begin{tabular}{lll}
\hline
Alternative & Density & Notation \\
\hline
Weibull & $\theta x^{\theta - 1} \exp \left( - x^{\theta} \right) $ & $W \left( \theta \right)$ \\
Lognormal & $\exp \left\{ - \frac{1}{2} \left( \frac{\log \left( x \right)}{\theta} \right)^{2} \right\} \Big/ \left\{ \theta x \sqrt{2 \pi} \right\}$ & $LN \left( \theta \right)$ \\
Gamma & $\frac{1}{\Gamma \left( \theta \right)} \, x^{\theta - 1} \exp \left( - x \right)$ & $\Gamma \left( \theta \right)$ \\
Chi-squared & $\frac{1}{2^{\theta / 2} \, \Gamma \left( \theta / 2 \right)} \, x^{\theta / 2 - 1} \exp \left( - \frac{x}{2} \right)$ & $\chi^{2} \left( \theta \right)$ \\
Dhillon & $\frac{\theta + 1}{x + 1} \, \exp \left\{ - \big( \log \left( x + 1 \right) \big)^{\theta + 1} \right\} \big( \log \left( x + 1 \right) \big)^{\theta}$ & $DH \left( \theta \right)$ \\\hline
\end{tabular}
\caption{Alternative distributions in consideration.}
\label{Table alternative distributions}
\end{table}

Since the null distribution of the test statistic depends on the unknown value of the shape parameter, it is our intention to use a parametric bootstrap, as explained in Section \ref{Section bootstrap procedure}, to calculate critical values for the test statistics in consideration. Given that the parametric bootstrap is computationally demanding, we employ the \emph{warp-speed} method proposed by \cite{GPW:2013} to approximate the power of all the considered tests. Denote the number of Monte Carlo replications by $MC$, and recall that this method capitalizes on the repetition inherent in the Monte Carlo simulation to produce bootstrap replications, rather than relying on a separate 'bootstrap loop'. The procedure can be summarized as follows.

\begin{enumerate}
\item Obtain a sample $X_{1}, \dots, X_{n}$ from a distribution, say $F$, and estimate $\mu$ and $\lambda$ by $\widehat{\mu}_{n}$ and $\widehat{\lambda}_{n}$, respectively. Also, let $\widehat{\varphi}_{n} = \widehat{\lambda }_{n} / \widehat{\mu}_{n}$.

\item Obtain the scaled data $Y_{n, j} = X_{j} / \widehat{\mu}_{n}$ for each $j = 1, \dots, n$, and calculate the test statistic, \linebreak$S = S(Y_{n, 1}, \dots, Y_{n, n})$, say.

\item Generate a bootstrap sample $X_{1}^{\ast}, \dots, X_{n}^{\ast}$ by independently sampling from $\mbox{IG}(1, \widehat{\varphi }_{n})$. Also calculate \linebreak$\widehat{\mu}_{n}^{\ast} = \widehat{\mu}_{n}(X_{1}^{\ast}, \dots, X_{n}^{\ast})$ and $\widehat{\lambda }_{n}^{\ast} = \widehat{\lambda}_{n}(X_{1}^{\ast}, \dots, X_{n}^{\ast})$.

\item Scale the values in the bootstrap sample using $Y_{n, j}^{\ast} = X_{j}^{\ast} / \widehat{\mu}_{n}^{\ast}$, $j = 1, \dots, n$, and determine the value of the test statistic for these scaled bootstrap values, that is, $S^{\ast} = S(Y_{n, 1}^{\ast}, \dots, Y_{n, n}^{\ast})$.

\item Repeat steps 1 to 4 $MC$ times to obtain $S_{1}, \dots, S_{MC}$ and $S_{1}^{\ast}, \dots , S_{MC}^{\ast}$, where $S_{m}$ denotes the value of the test statistic calculated from the $m^{\text{th}}$ scaled sample data generated in step 1, and $S_{m}^{\ast}$ denotes the value of the bootstrap test statistic calculated from the single scaled bootstrap sample obtained in the $m^{\text{th}}$ iteration of the Monte Carlo simulation.

\item To obtain the power approximation, reject the null hypothesis for the $i^{\text{th}}$ sample whenever $S_{i} > S_{\left( \left \lfloor MC \cdot (1 - \alpha) \right \rfloor \right)}^{\ast}$, $i = 1, \dots, MC$, where $S_{(1)}^{\ast} \leq \dotso \leq S_{(MC)}^{\ast}$ are the ordered values of the statistics obtained from the bootstrap samples and $\lfloor\cdot\rfloor$ denotes the floor function.
\end{enumerate}

The warp-speed methodology described above is a computationally efficient alternative to the classical parametric bootstrap procedure (with a separate 'bootstrap loop') usually employed in power calculations in the presence of a shape parameter. The latter method is detailed and implemented in Section 5 of \cite{HK:2002}. Interestingly, extensive Monte Carlo simulations indicate that the power estimates obtained using the two bootstrap methods provide almost identical results for all but two of the test statistics used. The tests for which the power estimates obtained using the two methods differ are those proposed in \cite{HK:2002}. For these tests, the classical parametric bootstrap approach typically provides power estimates that are higher than those obtained using the warp-speed methodology. The differences observed between the two sets of estimated powers are ascribed to the different ways in which the numerical difficulties pointed out in Section \ref{Subsection test Henze Klar} manifest in the two methods.

Tables \ref{Table power results n = 30} and \ref{Table power results n = 50} at the end of the paper show the estimated powers obtained by the warp-speed bootstrap methodology with $50~000$ replications for sample sizes $30$ and $50$. The entries show the percentage of samples for which the null hypothesis was rejected, rounded to the closest integer. The nominal significance level is set to $10\%$ throughout. In order to ease comparison, the highest power against each alternative distribution is printed in bold in the tables. We divide our new tests into four categories, depending on the weight function and estimation technique. More precisely, we implement the two weight functions $w_a$ and $\widetilde{w}_a$ as in Sections \ref{Section introduction} and \ref{Subsection computational form}, distinguishing the resulting tests by the '$~~ \widetilde{} ~~$'-notation, and use upper indices $ML$ and $MO$ to indicate the use of maximum likelihood or moment estimators, respectively. For each of the resulting four classes, we consider three different values of the tuning parameter, namely $a = 0.1, 1, 10$. Several other values were also investigated, but due to the remarkable insensitivity of the tests with regard to the tuning parameter, we present only these three values in the numerical results. The mentioned insensitivity is particularly noticeable for the tests that employ the moment estimators. \\

When comparing the power results presented in Tables \ref{Table power results n = 30} and \ref{Table power results n = 50}, some remarks are in order. Notice that each of the tests keeps the nominal level of $10\%$ closely when the null hypothesis is true. Observing the powers associated with the existing tests for the inverse Gaussian distribution, it is clear that the power of the test by \cite{VG:2015} is generally lower than the power of the remaining tests. Among the three classical procedures, the Anderson-Darling test performs best, while the $HK^{(2)}_{n,0}$ test proposed by \cite{HK:2002} produces the highest power among the more recent tests. \\

Turning our attention to a global comparison between the tests, we see that our newly proposed method outperforms the existing tests uniformly for sample size $n = 30$, and when $n = 50$, the new tests produce higher powers in $19$ out of $20$ cases. The only exception is the $LN\left( 3 \right)$ distribution. Interestingly, the tests using the moment estimators outperform those based on the maximum likelihood estimates. Although the choice of the weight function seems to have a smaller influence on the power than the estimation method, the test statistics which employ $w_a(t) = \exp(- a t)$ as a weight outperform those that use $\widetilde{w}_a(t) = \exp(- a t^2)$. Based on the numerical results, we recommend for practical applications the statistic $T_{n, 10}^{MO}$.

\section{Practical application}
\label{Section practical applications}

We apply all tests from the simulation study in Section \ref{Section monte carlo} to two real-world data examples. The first data set is from \cite{vA:1964}. It was also analyzed by \cite{GDAM:1997} and \cite{HK:2002} with regard to the inverse Gaussianity hypothesis stated by \cite{CF:1977}. The data is recalled in Table \ref{Table repair times for airborne transceivers}, where $n = 46$ active repair times (in hours) for an airborne transceiver are provided. Table \ref{Table analysis of the first data example} shows the calculated value of each test statistic as well as the associated $p$-value. Each of the $p$-values are calculated using the classical parametric bootstrap approach used in Section 5 of \cite{HK:2002}. Observing the $p$-values, none of the tests rejects the null hypothesis at a nominal significance level of 10\%. \\

\begin{table}%
\small
\centering\begin{tabular}{cccccccccccc}
\hline
0.2 & 0.3 & 0.5 & 0.5 & 0.5 & 0.5 & 0.6 & 0.6 & 0.7 & 0.7 & 0.7 & 0.8 \\
0.8 & 1.0 & 1.0 & 1.0 & 1.0 & 1.1 & 1.3 & 1.5 & 1.5 & 1.5 & 1.5 & 2.0 \\
2.0 & 2.2 & 2.5 & 2.7 & 3.0 & 3.0 & 3.3 & 3.3 & 4.0 & 4.0 & 4.5 & 4.7 \\
5.0 & 5.4 & 5.4 & 7.0 & 7.5 & 8.8 & 9.0 & 10.3 & 22.0 & 24.5\\
\hline
\end{tabular}
\caption{Repair times (in hours) for airborne transceivers.}
\label{Table repair times for airborne transceivers}
\end{table}

\begin{table}%
\small
\centering\begin{tabular}{|ccc|ccc|}
\hline
Test & Test statistic & $p$-value & Test & Test statistic & $p$-value \\
\hline
$KS$ & 0.0682 & 0.9040  & $T^{MO}_{46, 0.1}$ & 4.0310 & 0.6423  \\
$CM$ & 0.0327 & 0.8707 & $T^{MO}_{46, 1}$ & 0.4870 & 0.6643  \\
$AD$ & 0.2195 & 0.8826  & $T^{MO}_{46, 10}$ & 0.0223 & 0.6641  \\
$HK^{(1)}_{46, 0}$ & 0.0137 & 0.9409  & $\widetilde{T}^{ML}_{46, 0.1}$ & 0.0618 & 0.9445  \\
$HK^{(2)}_{46, 0}$ & 0.0028 & 0.9608  & $\widetilde{T}^{ML}_{46, 1}$ & 0.0320 & 0.8557  \\
$VG_{46}$ & 0.5770 & 0.7115  & $\widetilde{T}^{ML}_{46, 10}$ & 0.0101 & 0.7923  \\
$T^{ML}_{46, 0.1}$ & 0.0949 & 0.9199  & $\widetilde{T}^{MO}_{46, 0.1}$ & 1.3230 & 0.6479 \\
$T^{ML}_{46, 1}$ & 0.0298 & 0.9073  & $\widetilde{T}^{MO}_{46, 1}$ & 0.4588 & 0.6577  \\
$T^{ML}_{46, 10}$  & 0.0020 & 0.8436  & $\widetilde{T}^{MO}_{46, 10}$ & 0.1227 & 0.6399  \\
\hline
\end{tabular}
\caption{Analysis of the data example from Table \ref{Table repair times for airborne transceivers}.}
\label{Table analysis of the first data example}
\end{table}

We now turn our attention to a second example, with data taken from \cite{AT:1975} as analyzed by \cite{OR:1992} and \cite{HK:2002}, where the inverse Gaussian distribution as an underlying model was again suggested by \cite{FC:1978}. The $n = 25$ recorded measurements correspond to precipitation (in inches) at Jug Bridge, Maryland. Table \ref{Table precipitation at Jug Bridge} provides the data itself, while Table \ref{Table analysis of the second data example} shows the values of the different test statistics as well as the associated $p$-value for each test. In contrast to the previous example, four of the tests reject the null hypothesis at a nominal significance level of $10\%$. This casts some doubt as to whether or not the data in question was realized from an inverse Gaussian law. Note that the $p$-values associated with the newly proposed tests are substantially lower than those associated with the existing tests.

\begin{table}
\small
\centering\begin{tabular}{ccccccccccccc}
\hline
1.01 & 1.11 & 1.13 & 1.15 & 1.16 & 1.17 & 1.17 & 1.20 & 1.52 & 1.54 & 1.54 & 1.57 & 1.64 \\
1.73 & 1.79 & 2.09 & 2.09 & 2.57 & 2.75 & 2.93 & 3.19 & 3.54 & 3.57 & 5.11 & 5.62 \\
\hline
\end{tabular}
\caption{Precipitation (in inches) at Jug Bridge, Maryland.}
\label{Table precipitation at Jug Bridge}
\end{table}

\begin{table}
\small
\centering\begin{tabular}{|ccc|ccc|}
\hline
Test & Test statistic & $p$-value & Test & Test statistic & $p$-value \\
\hline
$KS$ & 0.0679 & 0.9127 & $T^{MO}_{25, 0.1}$ & 1.9691 & 0.4380 \\
$CM$ & 0.0307 & 0.8944 & $T^{MO}_{25, 1}$ & 0.2903 & 0.4357 \\
$AD$ & 0.2110 & 0.8968 & $T^{MO}_{25, 10}$ & 0.0123 & 0.3350 \\
$HK^{(1)}_{25, 0}$ & 0.0121 & 0.9482 & $\widetilde{T}^{ML}_{25, 0.1}$ & 0.3203 & 0.1080 \\
$HK^{(2)}_{25, 0}$ & 0.0025 & 0.9630 & $\widetilde{T}^{ML}_{25, 1}$ & 0.1796 & 0.0441 \\
$VG_{25}$ & 0.4314 & 0.7209 & $\widetilde{T}^{ML}_{25, 10}$ & 0.0195 & 0.0584 \\
$T^{ML}_{25, 0.1}$ & 0.3216 & 0.1068 & $\widetilde{T}^{MO}_{25, 0.1}$ & 0.6804 & 0.4321 \\
$T^{ML}_{25, 1}$ & 0.1581 & 0.0632 & $\widetilde{T}^{MO}_{25, 1}$ & 0.3005 & 0.4356 \\
$T^{ML}_{25, 10}$  & 0.0029 & 0.0720 & $\widetilde{T}^{MO}_{25, 10}$ & 0.0825 & 0.3315 \\
\hline
\end{tabular}
\caption{Analysis of the data example from Table \ref{Table precipitation at Jug Bridge}.}
\label{Table analysis of the second data example}
\end{table}

\section{Conclusions}
\label{Section conclusions}
Starting with the characterization of the inverse Gaussian distribution which results as a special case of the general identities provided by \cite{BE:2019:1}, we set out to construct a goodness-of-fit test for inverse Gaussianity. The test is based on a weighted $L^2$-statistic, a class of statistics which is extensively studied and extraordinarily well understood. After introducing our new testing procedure, we recalled the maximum likelihood and moment estimators for the parameters of the inverse Gaussian distribution, focusing on asymptotic expansions that were needed in the subsequent section to derive the limit null distribution. We briefly summarized the behavior of the novel statistic under contiguous alternatives to the hypothesis, and then turned to proving consistency. The main point of Section \ref{Section bootstrap procedure} was the proof of the consistency of the test in the bootstrap setting, thus taking into account precisely how the test is implemented in practice. In this endeavor, we also argued that the test based on the bootstrap critical values keeps its nominal level asymptotically. Monte Carlo simulations further indicate that the level is also kept for moderate, finite sample sizes ($n = 30$). Moreover, the power simulation study puts our method in strong favor, as it beats all competing tests virtually uniformly. In particular, the new test is more powerful than the other quite recent tests we considered and it also outperforms the classical empirical distribution function based tests, except in one instance of all alternative distributions where the Cram\'{e}r-von Mises Anderson-Darling test have higher power. We concluded the paper by considering a data example from reliability engineering and one data set from meteorology, for which we tested the hypothesis that the data sets were realized from some inverse Gaussian law.

Finally, we want to point out some interesting open questions for further research. In view of the data driven method of choosing a tuning parameter proposed by \cite{AS:2015} and \cite{T:2019} for location scale families, it would be beneficial for the power of the tests to find an optimal choice of $a$, even though our the procedures are not very sensitive to the choice of different tuning parameters. Another interesting question is to find characterizations similar to Theorem \ref{Theorem characterization} for larger classes of distributions, i.e. supersets of $\mathcal{IG}$, like the generalized hyperbolic distributions, and apply a similar methodology as the presented one to find powerful goodness-of-fit tests. For initial results on Stein characterisations of the generalized hyperbolic distribution, see \cite{G:2017}.

\bibliography{lit-IG}
\bibliographystyle{apalike}

\begin{landscape}
\begin{table}
\small
\centering\begin{tabular}{|l|p{0.4cm}p{0.4cm}p{0.4cm}ccc|ccc|ccc|ccc|ccc|}
\hline
Distribution & $KS$ & $CM$ & $AD$ & $HK^{(1)}_{n, 0}$ & $HK^{(2)}_{n, 0}$ & $VG_n$ & $T^{ML}_{n, 0.1}$ & $T^{ML}_{n, 1}$ & $T^{ML}_{n, 10}$ & $T^{MO}_{n, 0.1}$ & $T^{MO}_{n, 1}$ & $T^{MO}_{n, 10}$ & $\widetilde{T}^{ML}_{n, 0.1}$ & $\widetilde{T}^{ML}_{n, 1}$ & $\widetilde{T}^{ML}_{n, 10}$ & $\widetilde{T}^{MO}_{n, 0.1}$ & $\widetilde{T}^{MO}_{n, 1}$ & $\widetilde{T}^{MO}_{n, 10}$ \\
\hline
$IG(1)$ & 10 & 10 & 10 & 10 & 10 & 10 & 10 & 10 & 10 & 12 & 12 & 13 & 10 & 10 & 11 & 12 & 12 & 12\\
$IG(5)$ & 10 & 10 & 10 & 10 & 10 & 10 & 11 & 10 & 10 & 11 & 11 & 11 & 11 & 10 & 10 & 11 & 11 & 11\\
$IG(10)$ & 10 & 10 & 10 & 10 & 10 & 10 & 10 & 11 & 10 & 11 & 11 & 11 & 10 & 10 & 10 & 10 & 11 & 11\\
$IG(20)$ & 10 & 10 & 10 & 10 & 10 & 11 & 10 & 10 & 11 & 10 & 10 & 10 & 11 & 11 & 11 & 10 & 11 & 10\\
$W(1)$ & 80 & 83 & 84 & 61 & 68 & 30 & 82 & 86 & 85 & 94 & 93 & \textbf{95} & 85 & 84 & 85 & 94 & 93 & 94\\
$W(1.2)$ & 73 & 76 & 77 & 58 & 64 & 36 & 77 & 80 & 78 & 91 & 90 & \textbf{92} & 79 & 79 & 79 & 90 & 90 & 91\\
$W(1.6)$ & 62 & 67 & 69 & 54 & 58 & 42 & 69 & 73 & 69 & 86 & 85 & \textbf{87} & 71 & 72 & 70 & 86 & 85 & 86\\
$W(2)$ & 55 & 61 & 63 & 52 & 54 & 44 & 65 & 67 & 64 & 82 & 82 & \textbf{83} & 66 & 67 & 66 & 82 & 82 & \textbf{83}\\
$W(3)$ & 47 & 53 & 56 & 48 & 49 & 44 & 61 & 63 & 59 & \textbf{76} & 75 & \textbf{76} & 61 & 63 & 60 & 75 & \textbf{76} & \textbf{76}\\
$LN(0.6)$ & 14 & 15 & 16 & 17 & 18 & 14 & 15 & 16 & 20 & 20 & 19 & \textbf{22} & 15 & 16 & 19 & 19 & 19 & 21\\
$LN(1)$ & 21 & 22 & 23 & 22 & 25 & 15 & 17 & 20 & 28 & 27 & 26 & \textbf{29} & 18 & 20 & 24 & 26 & 25 & 27\\
$LN(1.4)$ & 28 & 32 & 32 & 28 & 33 & 16 & 21 & 25 & 36 & 35 & 35 & \textbf{38} & 22 & 25 & 31 & 35 & 34 & 35\\
$LN(2)$ & 42 & 46 & 46 & 37 & 44 & 17 & 26 & 34 & 48 & 49 & 48 & \textbf{52} & 28 & 33 & 41 & 49 & 48 & 49\\
$LN(3)$ & 63 & 67 & 67 & 50 & 61 & 17 & 33 & 44 & 65 & 71 & 70 & \textbf{73} & 36 & 44 & 54 & 71 & 69 & 71\\
$\Gamma(1)$ & 81 & 84 & 84 & 61 & 68 & 30 & 82 & 86 & 85 & 94 & 93 & \textbf{95} & 85 & 85 & 85 & 94 & 93 & 94\\
$\Gamma(1.5)$ & 61 & 65 & 66 & 53 & 57 & 36 & 63 & 68 & 68 & 83 & 82 & \textbf{85} & 66 & 67 & 68 & 83 & 82 & 83\\
$\Gamma(2)$ & 47 & 52 & 53 & 47 & 50 & 35 & 51 & 55 & 57 & 73 & 71 & \textbf{75} & 53 & 55 & 57 & 72 & 71 & 74\\
$\Gamma(2.5)$ & 39 & 42 & 44 & 41 & 44 & 31 & 43 & 47 & 50 & 65 & 64 & \textbf{67} & 44 & 47 & 50 & 64 & 63 & 66\\
$\chi^2(3)$ & 60 & 65 & 66 & 53 & 57 & 36 & 63 & 68 & 68 & 83 & 82 & \textbf{85} & 65 & 67 & 69 & 83 & 82 & 83\\
$\chi^2(5)$ & 38 & 43 & 44 & 41 & 44 & 32 & 43 & 47 & 50 & 64 & 63 & \textbf{67} & 44 & 47 & 50 & 64 & 63 & 66\\
$\chi^2(10)$ & 21 & 23 & 24 & 27 & 28 & 21 & 26 & 27 & 32 & 41 & 40 & \textbf{43} & 26 & 29 & 32 & 40 & 39 & 42\\
$DH(1)$ & 39 & 42 & 43 & 39 & 43 & 27 & 37 & 43 & 49 & 56 & 55 & \textbf{59} & 39 & 43 & 47 & 55 & 54 & 57\\
$DH(1.5)$ & 35 & 39 & 39 & 38 & 41 & 29 & 37 & 42 & 47 & 57 & 57 & \textbf{61} & 39 & 42 & 46 & 57 & 56 & 58\\
$DH(2)$ & 33 & 37 & 38 & 38 & 40 & 29 & 39 & 42 & 45 & 58 & 57 & \textbf{60} & 39 & 42 & 46 & 57 & 56 & 59\\
\hline
\end{tabular}
\caption{Approximated powers for sample size $n = 30$.}
\label{Table power results n = 30}
\end{table}

\end{landscape}


\begin{landscape}
\begin{table}%
\small
\centering\begin{tabular}{|l|p{0.4cm}p{0.4cm}p{0.4cm}ccc|ccc|ccc|ccc|ccc|}
\hline
Distribution & $KS$ & $CM$ & $AD$ & $HK^{(1)}_{n, 0}$ & $HK^{(2)}_{n, 0}$ & $VG_n$ & $T^{ML}_{n, 0.1}$ & $T^{ML}_{n, 1}$ & $T^{ML}_{n, 10}$ & $T^{MO}_{n, 0.1}$ & $T^{MO}_{n, 1}$ & $T^{MO}_{n, 10}$ & $\widetilde{T}^{ML}_{n, 0.1}$ & $\widetilde{T}^{ML}_{n, 1}$ & $\widetilde{T}^{ML}_{n, 10}$ & $\widetilde{T}^{MO}_{n, 0.1}$ & $\widetilde{T}^{MO}_{n, 1}$ & $\widetilde{T}^{MO}_{n, 10}$ \\
\hline
$IG(1)$ & 10 & 10 & 10 & 10 & 10 & 10 & 10 & 10 & 11 & 11 & 11 & 11 & 10 & 10 & 10 & 11 & 11 & 11 \\
$IG(5)$ & 10 & 10 & 10 & 10 & 10 & 10 & 10 & 10 & 10 & 11 & 10 & 11 & 11 & 10 & 10 & 10 & 10 & 10 \\
$IG(10)$ & 10 & 10 & 10 & 10 & 10 & 10 & 10 & 10 & 10 & 11 & 10 & 10 & 11 & 10 & 10 & 10 & 10 & 10 \\
$IG(20)$ & 10 & 10 & 10 & 10 & 10 & 10 & 10 & 10 & 10 & 11 & 10 & 10 & 11 & 10 & 10 & 10 & 10 & 10 \\
$W(1)$ & 94 & 96 & 96 & 79 & 85 & 36 & 96 & 97 & 96 & 98 & 98 & \textbf{99} & 97 & 96 & 96 & 98 & 98 & \textbf{99} \\
$W(1.2)$ & 89 & 92 & 92 & 77 & 82 & 45 & 93 & 95 & 92 & \textbf{98} & 97 & \textbf{98} & 94 & 94 & 93 & 97 & 97 & \textbf{98} \\
$W(1.6)$ & 81 & 85 & 86 & 74 & 77 & 57 & 88 & 90 & 86 & \textbf{96} & 95 & \textbf{96} & 89 & 89 & 87 & \textbf{96} & 95 & \textbf{96} \\
$W(2)$ & 74 & 80 & 82 & 71 & 74 & 62 & 85 & 86 & 82 & \textbf{94} & \textbf{94} & \textbf{94} & 86 & 86 & 84 & \textbf{94} & \textbf{94} & \textbf{94} \\
$W(3)$ & 65 & 73 & 76 & 68 & 69 & 63 & 82 & 82 & 78 & \textbf{91} & 90 & \textbf{91} & 82 & 82 & 79 & 90 & 90 & 90 \\
$LN(0.6)$ & 16 & 18 & 18 & 20 & 21 & 15 & 17 & 18 & \textbf{23} & 21 & 20 & \textbf{23} & 17 & 18 & 22 & 20 & 20 & 21 \\
$LN(1)$ & 26 & 29 & 30 & 29 & 33 & 18 & 21 & 26 & \textbf{35} & 28 & 27 & 31 & 22 & 26 & 31 & 28 & 27 & 28 \\
$LN(1.4)$ & 38 & 42 & 43 & 39 & 44 & 19 & 27 & 35 & \textbf{47} & 36 & 35 & 40 & 29 & 36 & 41 & 36 & 34 & 36 \\
$LN(2)$ & 57 & 62 & 62 & 52 & 60 & 18 & 36 & 49 & \textbf{63} & 49 & 48 & 53 & 41 & 49 & 56 & 49 & 48 & 50 \\
$LN(3)$ & 81 & \textbf{84} & \textbf{84} & 68 & 78 & 18 & 48 & 66 & 82 & 72 & 71 & 75 & 56 & 65 & 74 & 71 & 71 & 72 \\
$\Gamma(1)$ & 94 & 95 & 95 & 78 & 84 & 35 & 96 & 97 & 96 & 98 & 98 & \textbf{99} & 97 & 96 & 96 & 98 & 98 & \textbf{99} \\
$\Gamma(1.5)$ & 79 & 83 & 84 & 72 & 76 & 48 & 83 & 87 & 85 & 93 & 93 & \textbf{94} & 86 & 85 & 86 & 93 & 92 & \textbf{94} \\
$\Gamma(2)$ & 65 & 71 & 72 & 64 & 68 & 48 & 70 & 75 & 75 & 85 & 85 & \textbf{88} & 73 & 74 & 75 & 85 & 84 & 87 \\
$\Gamma(2.5)$ & 53 & 59 & 60 & 58 & 60 & 45 & 59 & 64 & 67 & 77 & 76 & \textbf{80} & 61 & 64 & 67 & 77 & 76 & 79 \\
$\chi^2(3)$ & 79 & 83 & 84 & 72 & 76 & 49 & 83 & 87 & 85 & 93 & 93 & \textbf{94} & 86 & 85 & 86 & 93 & 92 & \textbf{94} \\
$\chi^2(5)$ & 53 & 59 & 60 & 58 & 60 & 45 & 59 & 64 & 67 & 77 & 76 & \textbf{80} & 61 & 64 & 67 & 77 & 76 & 79 \\
$\chi^2(10)$ & 27 & 31 & 32 & 38 & 39 & 30 & 36 & 39 & 43 & 51 & 50 & \textbf{54} & 36 & 40 & 44 & 50 & 49 & 53 \\
$DH(1)$ & 53 & 58 & 59 & 54 & 59 & 34 & 51 & 59 & 66 & 66 & 64 & \textbf{70} & 55 & 59 & 63 & 65 & 63 & 67 \\
$DH(1.5)$ & 48 & 53 & 54 & 53 & 56 & 38 & 52 & 58 & 63 & 69 & 67 & \textbf{73} & 55 & 58 & 61 & 68 & 67 & 71 \\
$DH(2)$ & 46 & 51 & 53 & 53 & 56 & 42 & 54 & 58 & 61 & 71 & 70 & \textbf{74} & 56 & 58 & 61 & 71 & 69 & 73 \\
\hline
\end{tabular}
\caption{Approximated powers for sample size $n = 50$.}
\label{Table power results n = 50}
\end{table}

\end{landscape}
\end{document}